# An Introduction to Causal Inference Methods with Multi-omics Data

Minhao Yao[1], Zhonghua Liu[2*]

[1] Department of Statistics and Actuarial Science, University of Hong Kong, Hong Kong

[2] Department of Biostatistics, Columbia University, New York, NY, USA

* Correspondence to: Zhonghua Liu (zl2509@cumc.columbia.edu).

## Abstract

Omics biomarkers play a pivotal role in personalized medicine by providing molecular-level insights into the etiology of diseases, guiding precise diagnostics, and facilitating targeted therapeutic interventions. Recent advancements in omics technologies have resulted in an increasing abundance of multimodal omics data, providing unprecedented opportunities for identifying novel omics biomarkers for human diseases. Mendelian randomization (MR) is a practically useful causal inference method that uses genetic variants as instrumental variables (IVs) to infer causal relationships between omics biomarkers and complex traits/diseases by removing hidden confounding bias. In this article, we first present current challenges in performing MR analysis with omics data, and then describe four MR methods for analyzing multi-omics data including epigenomics, transcriptomics, proteomics, and metabolomics data, all executable within the R software environment.

## 1. Introduction:

Mendelian randomization (MR) is a powerful causal inference method that utilizes genetic variants, usually single-nucleotide polymorphisms (SNPs), as instrumental variables (IVs) to assess the causal effect of a modifiable exposure on an outcome of interest in observational studies (Davey Smith & Ebrahim, 2003; Lawlor, Harbord, Sterne, Timpson, & Davey Smith, 2008; Davey Smith & Hemani, 2014; Sanderson et al., 2022). To provide reliable scientific findings, conventional MR method requires that the genetic variants included in the analysis are valid IVs, that is, each genetic variant must satisfy the following three core IV assumptions (Didelez & Sheehan, 2007; Lawlor et al., 2008; Sanderson et al., 2022):

(i) Relevance: the genetic variant is associated with the exposure;

(ii) Independence: the genetic variant is not associated with any unmeasured confounder for the exposure-outcome relationship; and

(iii) Exclusion restriction: the genetic variant influences the outcome only through the exposure in view.

If all the selected genetic variants used in the MR analysis are valid IVs, then the inverse-variance weighted (IVW) method (Lawlor et al., 2008) can be applied to obtain a consistent estimate of the causal effect. In the literature, many MR methods have been proposed to address the potential violations of the three core IV assumptions (Bowden, Davey Smith, & Burgess, 2015; Bowden, Davey Smith, Haycock, & Burgess, 2016; Bowden et al., 2017; Xu, Fung, & Liu, 2021; Xu, Wang, Fung, & Liu, 2023; Yao, Guo, & Liu, 2023). For a more comprehensive review, please see Sanderson et al. (2022).

Large-scale multi-omics data are increasingly available, for example, epigenomics, transcriptomics, proteomics, and metabolomics data (Hasin, Seldin, & Lusis, 2017). Multi-omics data play a pivotal role in providing a holistic view of the complex biological systems, allowing researchers to uncover relationships, patterns, and potential intervention targets across multiple biological components. Therefore, to comprehensively investigate the underlying biological mechanisms for the phenotypic outcome of interest, it is promising to perform omics MR (xMR, a term firstly coined in Yao et al. (2023)) to assess the (total) average causal effects of omics biomarkers on complex traits and diseases, and to infer biological pathways (direct and indirect effects) in the causal mediation analysis framework by removing hidden confounding bias.

However, xMR analyses encounter notable challenges when applied to multi-omics data. First, the magnitude of effect sizes for genetic variants varies among different omics data types. For example, the effect sizes of protein quantitative trait loci (pQTLs) are observed to be generally smaller compared to those of expression quantitative trait loci (eQTLs), suggesting potential buffering at the protein level (Vandiedonck, 2018). Limited sample sizes further contribute to the challenge of replicating pQTLs, resulting in a scarcity of genetic instruments for xMR analysis in the context of pQTLs (Vandiedonck, 2018; Sun et al., 2023). Second, the high dimensionality of omics data complicates the task of discerning causal omics biomarkers (Berger, Peng, & Singh, 2013; Picard, Scott-Boyer, Bodein, Périn, & Droit, 2021). Third, intricate interdependencies among omics biomarkers add complexity to integrating multiple omics data types, hindering the identification of biological pathways influencing the phenotype of interest (e.g., genetic variants -> DNA methylation -> transcripts -> proteins -> metabolites -> diseases) (Subramanian, Verma, Kumar, Jere, & Anamika, 2020; Graw et al., 2021; Vahabi & Michailidis, 2022).

To address those challenges, a number of xMR methods have been developed specifically for detecting putatively causal omics biomarkers for complex traits and diseases (Relton & Davey Smith, 2012; Porcu et al., 2019; Zheng et al., 2020; Zuber, Colijn, Klaver, & Burgess, 2020). A comprehensive review of existing xMR methods is beyond the scope of this article. Here, we aim to provide a tutorial-style introduction to xMR methods for analyzing multi-omics data based on our prior experience.

## 2. Transcriptome-wide Mendelian Randomization (TWMR) Methods

Genome-wide association studies (GWAS) have discovered numerous genetic variants associated with thousands of complex traits and diseases(Wang, Barratt, Clayton, & Todd, 2005; Tam et al., 2019; Uffelmann et al., 2021). Many of those identified genetic variants linked to traits are also associated with gene expressions, known as expression quantitative trait loci (eQTLs), suggesting that eQTLs may serve as instrumental variables for analyzing the impact of gene expression levels on complex traits and diseases (Nica et al., 2010; Fehrmann et al., 2011; Hernandez et al., 2012). Furthermore, eQTLs frequently exhibit shared associations with multiple genes (Westra et al., 2013), thus it is desirable to consider multiple gene expressions as exposures simultaneously. Recent studies have developed practical methods for discovering causal genes from TWAS data (Barfield et al., 2018; Porcu et al., 2019; Gleason, Yang, & Chen, 2021; Zhao et al., 2024), and references therein. Notably, Porcu et al. (2019) proposed the Transcriptome-Wide Mendelian Randomization (TWMR) to identify gene expressions that are associated with an outcome in a multivariate-MR framework using summary statistics from GWAS and eQTL studies. The R package for TWMR is available at https://github.com/eleporcu/TWMR.

To run TWMR, we first need to prepare two input files:

1. An $n \times (k + 2)$ matrix where the first column consists of the rsid of $n$ SNPs, followed by $k$ columns containing the estimated effect sizes of $n$ SNPs on $k$ genes from eQTL studies, with the last row indicating the estimated effect size of $n$ SNPs on the outcome of interest from GWAS studies. This file should be named with the suffix ".matrix". For example, the example at https://github.com/eleporcu/TWMR provides a file named "ENSG00000002919.matrix", whose first five rows are given as follows:

**Table 1:** The first five rows of the file "ENSG00000002919.matrix".

| GENES | ENSG00000002919 | ENSG00000159202 | BETA_GWAS |
|---|---|---|---|
| rs221602 | -2.495E-02 | 0.000E+00 | 3.247E-03 |
| rs1317850 | 1.481E-01 | 0.000E+00 | -1.617E-04 |
| rs1468270 | -3.096E-01 | 0.000E+00 | 8.533E-03 |
| rs7350950 | 0.000E+00 | -4.463E-02 | 6.919E-03 |
| rs9897918 | -6.519E-02 | 0.000E+00 | -1.193E-05 |

2. An $n \times n$ matrix containing the linkage disequilibrium (LD) matrix of $n$ SNPs from reference panel, for example, the 1000 Genomes (Consortium, 2010). This file should be named with the suffix ".ld", and the preceding name must match the name of the effect size matrix file. For example, if the name of the effect size matrix file is "ENSG00000002919.matrix", then this LD matrix file should be named as "ENSG00000002919.ld".

With well-formatted input files, we can perform TWMR by running the R script "MR.R". Note that the numbers in the third and fourth rows should be replaced by the sample sizes of the GWAS study and eQTL study, respectively. Suppose the name of the effect size matrix file is "ENSG00000002919.matrix", then we can run "MR.R" script in the terminal as follows:

```
R < MR.R --no-save ENSG00000002919
```

This command will output one file with suffix ".alpha" that contains the following columns:

- gene: the name of the gene tested.
- alpha: the causal effect of the gene on the outcome estimated by TWMR.
- SE: the standard error of the estimated causal effect calculated by the delta method.
- P: the p-value of the estimated causal effect.
- Nsnps: the number of SNPs used.
- Ngene: the number of genes included.

## 2.1 Selection of genetic variants and genes for TWMR

TWMR uses the following procedure to select genetic variants for multivariate MR analysis:

- Step 1: choose one gene.
- Step 2: select independent and significant eQTLs for this gene by linkage disequilibrium (LD) clumping using plink software (Purcell et al., 2007) or TwoSampleMR R package (Hemani et al., 2018).
- Step 3: include all genes for which the selected SNPs in Step 2 are eQTLs.
- Step 4: include all SNPs that are eQTLs only for genes in Step 3.
- Step 5: perform LD clumping for all SNPs in Step 4 to keep only independent SNPs as IVs.

## 3. MR Method for Proteomics Data

Plasma proteins play crucial roles in various biological processes and serve as promising targets for drug discovery and drug repurposing (Imming, Sinning, & Meyer, 2006; Santos et al., 2017; Yao et al., 2023). Recent analyses of genetic associations with the plasma proteome have detected thousands of independent SNPs associated with proteins, which are referred to as protein quantitative trait loci (pQTLs), allowing for the application of MR to assess the causal effects of plasma proteins on complex traits and diseases(Emilsson et al., 2018; Sun et al., 2018). Based on the location on the chromosome relative to either side of the transcription starting site of one gene, the pQTLs can be categorized into the following two types:

1. *cis*-acting pQTLs: located near the encoding gene, for example, within a 500-kb window of the leading pQTL.
2. *trans*-acting pQTLs: located outside the 500-kb window of *cis*-acting pQTLs.

Since *cis*-acting pQTLs are nearer to the encoding genes, they have been employed in phenome-wide scans for potential drug targets (Millwood et al., 2018). In comparison, *trans*-acting pQTLs typically function through indirect mechanisms, and thus often exhibit more pleiotropic effects (Swerdlow et al., 2016). To utilize both *cis*-pQTLs and *trans*-pQTLs, we may conduct main MR analysis using *cis*-pQTLs as instruments (*cis*-only analysis), and then perform sensitivity analysis using both *cis*-pQTLs and *trans*-pQTLs as instruments (*cis* + *trans* analysis) and using *trans*-pQTLs only as instruments (*trans*-only analysis). Plasma proteomics data from UK Biobank Pharma Proteomics Project (UKB-PPP, Sun et al. (2023)) can be downloaded at http://ukb-ppp.gwas.eu/. Following Zheng et al. (2020), one can perform xMR analysis for proteomics data as follows:

- Step 1: select pQTLs. In this step, independent (after linkage disequilibrium (LD) clumping by plink software (Purcell et al., 2007) or TwoSampleMR R package (Hemani et al., 2018)) and significant (after Bonferroni correction (Bland & Altman, 1995)) pQTLs are selected as genetic instruments. In addition, pQTLs within the human major histocompatibility complex (MHC) region (chr6: from 26 Mb to 34 Mb, De Bakker et al. (2006)) should be excluded.

- Step 2: test validity of pQTLs. First, we count the number of proteins associated with each pQTL, and exclude pQTLs that are associated with five or more proteins, which are more likely to be highly pleiotropic, from downstream MR analysis. Second, we apply biological pathway information from Reactome database (https://reactome.org/ ) and protein-protein interaction information from STRING DB database (https://string-db.org/ ) to differentiate between vertical and horizontal pleiotropy. We retain pQTLs linked to multiple proteins and aligned with the same pathway or protein-protein interaction network as valid instruments. In addition to using the external biological information described above to test the validity of pQTLs, we can also apply the MR-SPI method (Yao et al., 2023) to automatically select a set of valid pQTLs for downstream analysis. The R package for MR-SPI is available at https://github.com/MinhaoYaooo/MR-SPI.

- Step 3: identify plasma proteins associated with phenotypes of interest through MR analysis. For proteins with one pQTL as instrument, we apply the Wald ratio method (Lawlor et al., 2008) to obtain an estimate of the causal effect. For proteins with two or more pQTLs as instruments, the inverse variance-weighted (IVW) method (Lawlor et al., 2008) is applied to obtain causal effect estimates. Additional MR methods like MR-Egger (Bowden et al., 2015) and the weighted median method (Bowden et al., 2016) can also be performed as sensitivity analyses for the robustness of the results. These MR methods can be implemented using the TwoSampleMR R package available at https://github.com/MRCIEU/TwoSampleMR. Specifically, we require the following two data frames to perform MR analysis: (1) *exp_dat* that contains the summary statistics for proteins, and (2) *out_dat* that contains the summary statistics for the outcome. Then we harmonize the above two data frames so that the effect alleles are aligned:

    *dat=harmonise_dat(exposure_dat=exp_dat, outcome_dat=out_dat)*

    Then we call the function `mr` in TwoSampleMR R package to perform MR analysis:

    *mr_results=mr(dat, method_list=c("mr_wald_ratio", "mr_ivw", "mr_egger_regression", "mr_weighted_median"))*

    The parameter *method_list* in function `mr` contains all the MR method we include in our MR analysis. We can call the function `mr_method_list` to obtain the list of all the available MR methods in TwoSampleMR R package.

- Step 4: repeat Step 3 for *cis*-only, *cis + trans*, and *trans*-only analyses, and then check whether the directions of causal effect estimates align consistently for each protein across the three analyses.

## 4. MR-BMA: MR Method for Metabolomics Data

Plasma metabolites serve as practically useful biomarkers of disease states because the measurement of metabolites is noninvasive (for example, proton nuclear magnetic resonance (1H-NMR) spectra of human serum (Brindle et al., 2002)), and that they are modifiable through lifestyle and diet (Enche Ady et al., 2017). Some metabolites (for example, lipid subfractions) are highly correlated both phenotypically and genetically, resulting in shared genetic variants among them (Zuber et al., 2020; Lord et al., 2021; Yin et al., 2022). To deal with this challenge, Zuber et al. (2020) proposed a two-sample multivariate MR approach based on Bayesian Model Averaging (MR-MBA) to detect putatively causal risk factors from potentially highly correlated metabolites, and the corresponding R package is available at https://github.com/verena-zuber/demo_AMD. Metabolomics data can be downloaded at https://metabolomips.org/gwas/.

We need the following two matrices in R to run MR-BMA:

1. betaX: an $n \times k$ matrix that contains the effect sizes of $n$ SNPs on $k$ metabolites from metabolomics data.

2. betaY: an $n \times 1$ matrix that contains the effect sizes of $n$ SNPs on the outcome of interest from GWAS data.

These two matrices are suggested to be weighted by the inverse of the standard errors of the outcome effect size estimates. We also require the following three objects to create the input object for MR-BMA:

1. rs: a vector of length $n$ that contains the rsid of $n$ SNPs.
2. rf: a vector of length $k$ that contains the names of $k$ metabolites.
3. outcome: a character indicating the name of the outcome.

Then we create an object of class `mvMRInput` as follows:

*input=new("mvMRInput", betaX = betaX, betaY = betaY, snps = rs, exposure = rf, outcome = outcome)*

With the object `input`, we can run MR-BMA by calling the function `summarymvMR_SSS`:

*BMA_output= summarymvMR_SSS(input,kmin=1,kmax=12, prior_prob=0.1, max_iter=100000)*

Note that the following parameters in function `summarymvMR_SSS` need to be adjusted according to the dataset under use:

1. kmin: the minimum model size, i.e., the minimum number of metabolites to be considered.
2. kmax: the maximum model size. For computational feasibility, it is better to set kmax<=12.
   a) If kmin=kmax, then an exhaustive search is applied.
   b) If kmin<kmax, then a stochastic search is applied.
3. prior_prob: the prior knowledge of the proportion of true causal biomarkers in the candidate metabolites. For example, if $k = 30$ and prior_prob = 0.1, then there are 3 expected causal biomarkers.
4. max_iter: the number of stochastic searches if kmin<kmax. For stable results, it is recommended to set max_iter to be no less than 100,000.

The output `BMA_output` produced by the function `summarymvMR_SSS` includes all the combinations of metabolites that satisfy the user-specified parameters. The best model with the highest posterior probability from MR_BMA can be obtained with the following codes:

*BMA_output@BestModel_Estimate      # causal effect estimates for each metabolite in the best model.*

*rf[BMA_output@BestModel_Estimate!=0]    # metabolites with non-zero causal effect estimates in the best model.*

We can call the function `sss.report.best.model` to obtain the next best models:

*sss.report.best.model(BMA_output, prior_sigma = 0.5, top = 10)*

The parameter `prior_sigma` is the prior knowledge of the standard deviation of the causal effects among the metabolites, and a large value of `prior_sigma` indicates a prior belief that the causal effects are strong. The parameter `top=10` means we wish to obtain the top 10 best models from MR-MBA.

We can also call the function `sss.report.mr.bma` to obtain the marginal inclusion probability (MIP) and model-average causal estimate (MACE) for each metabolite in the top best models:

*sss.report.mr.bma(BMA_output, top = 10)*

The MIP of a metabolite is calculated by the sum of the posterior probabilities over all models that include this metabolite, and thus a large value of MIP indicates that this metabolite is more likely to be a causal biomarker.

## 5. Two-step Epigenetic MR for Mediation Analysis

Epigenetic mechanism can modify gene activity without changing the DNA sequence (Weinhold, 2006; Dupont, Armant, & Brenner, 2009). DNA methylation, involving the addition or removal of a methyl group (CH3), primarily occurs at sites where cytosine bases are consecutive, serves as a crucial epigenetic marker regulating gene expression and chromatin organization (Suzuki & Bird, 2008). Recently, several methods have been proposed to investigate how DNA methylation mediates the effects of modifiable exposures on outcomes (Liu et al., 2022; Tian, Yao, Huang, & Liu, 2022; Clark-Boucher et al., 2023). However, the presence of potential confounding factors poses challenge to mediation analysis when analyzing DNA methylation data (Adalsteinsson et al., 2012; Houseman, Kim, Kelsey, & Wiencke, 2015). To deal with this challenge, Relton and Davey Smith (2012) proposes the following two-step epigenetic MR method for mediation analysis to uncover the causal relationships among exposure, DNA methylation and outcome of interest, as illustrated in Figure 1:

- Step 1: use genetic instruments to assess the causal effect of exposure on DNA methylation. For the validity of MR analysis, the selected genetic instruments in this step should not directly influence DNA methylation.

- Step 2: use genetic instruments to assess the causal effect of DNA methylation on outcome. In this step, SNPs located near the CpG site correlated with methylation levels can be considered as genetic instruments.

If the causal effect estimates in steps 1 and 2 are both statistically significant (the joint significance test), then the targeted DNA methylation CpG site can be considered as a mediator of the exposure on the outcome of interest.

## 6. Commentary

### *6.1 Background Information*

In this article, we provide a gentle tutorial-style introduction to four xMR methods for four different types of omics data including transcriptomics, metabolomics, proteomics, and epigenomics data. Notably, among the above four xMR methods, TWMR and MR-BMA belong to the multivariable MR approaches that allow for multiple exposures simultaneously in the analysis. For robust MR analysis, the selection of genetic instruments is critical. As noted in Zuber et al. (2020), the IV selection criteria of multivariable MR methods are slightly different from the univariable MR methods. First, since multiple exposures are included in the multivariable MR methods, the core assumption (iii) is more likely to be satisfied given a set of genetic variants due to the measured pleiotropy (Zuber et al., 2020). Second, in multivariable MR methods, not every genetic variant need to be associated with all the risk factors. Indeed, the genetic variants included in a multivariable MR method should jointly satisfy the following two criteria (Zuber et al., 2020): (C1): each genetic variant must be strongly associated with at least one exposure; and (C2): the genetic associations of any exposure cannot be the linear combination of genetic associations of the other exposures. More future work is still needed to systematically select a set of candidate genetic instruments for robust xMR analysis with omics data.

Given the high dimensionality of omics data, a multitude of potential biomarkers are often considered as candidate exposures in xMR analyses. While multivariable MR methods like TWMR and MR-BMA can handle multiple omics biomarkers concurrently, their limitations arise when dealing with the entirety of available biomarkers simultaneously. For instance, TWMR only focuses on genes with cis-eQTLs, and MR-BMA faces computational challenges with an extensive list of candidate metabolites. On the other hand, univariable MR methods conduct MR analyses separately for each omics biomarker, followed by multiple comparison procedure (e.g., Bonferroni correction) to detect significant associations. A major challenge is to develop more efficient approaches that can incorporate all candidate multi-omics biomarkers simultaneously within a single model framework to gain a holistic view of the biological functions and pathways of multi-omics biomarkers.

## *6.2 Critical Parameters*

1. R software version: the above MR methods can be implemented in both R version 3.x and R version 4.x.

2. Linkage disequilibrium (LD) clumping parameters: since the MR methods introduced in this article require independent genetic instruments, LD clumping for candidate SNPs should be performed before the downstream analysis. LD, typically denoted as $r^2$, is a measure of the correlation between two SNPs, and a large value of $r^2$ suggests a high correlation between two SNPs. LD clumping typically requires four parameters (*clump_kb, clump_r2, clump_p1,* and *clump_p2*) and one file of pre-calculated LD matrix as reference panel. The LD clumping contains the following steps: (1) select all SNPs reaching significance at *clump_p1* threshold and designate them as index SNPs if they have not already been clumped; (2) for each index SNP, select all other SNPs within distance *clump_kb* that reach significance at *clump_p2* threshold and are in LD ($r^2 <$ *clump_r2*) with the index SNP; (3) select the most significant SNP as the representative for those SNPs which are clumped together.

## *6.3 Troubleshooting*

Some common problems and the possible solutions are summarized in Table 2.

**Table 2**: Troubleshooting guide for common issues in MR methods for multi-omics data.

| Problem | Possible cause | Possible solution |
| --- | --- | --- |
| *NA* values in the output of TWMR | The dimensions of input files are not matched | Check whether the number of SNPs in the file with the suffix ".matrix" equals the number of rows/columns of the LD matrix file with the suffix ".ld" |
| Running time of MR-BMA is long | Too many candidate models need to be considered | Reduce the maximum model size (kmax) to a value below 12 and set kmin<kmax to use stochastic search |
| No output for MR-Egger/ weighted median method | The number of candidate SNPs are less than 3 | Increase the LD clumping threshold and/or p-value threshold to include more SNPs |

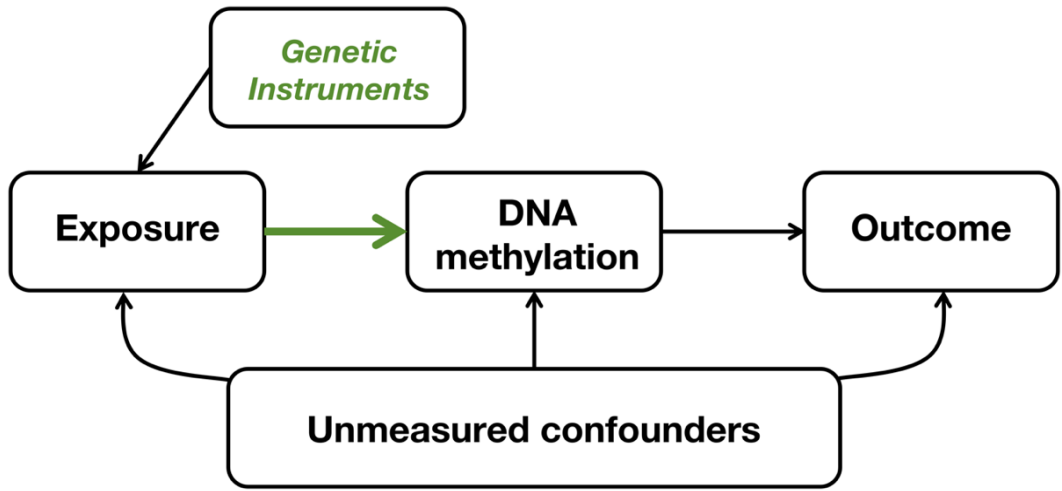

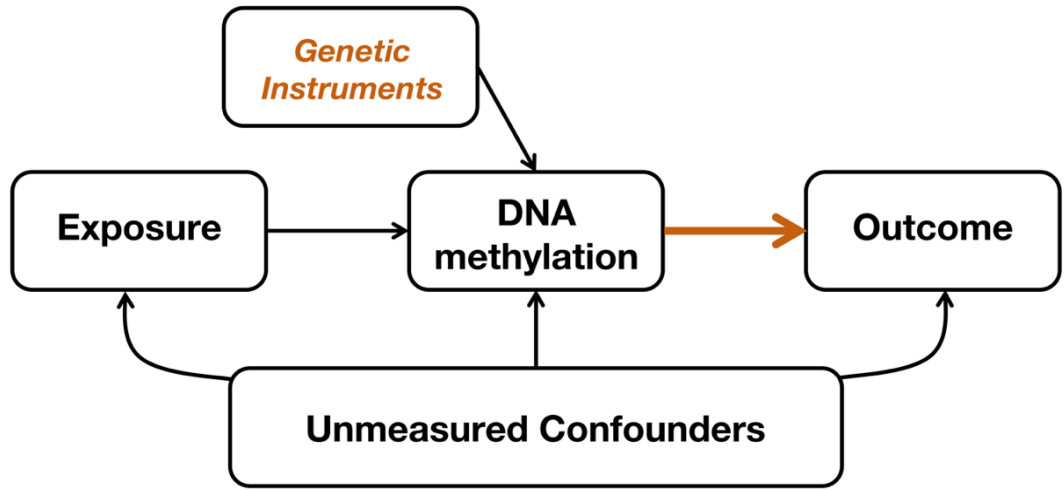

**Figure 1.** Two-step epigenetic MR for mediation analysis. The first step uses genetic instruments to assess the causal effect of the exposure on DNA methylation. The second step uses genetic instruments (possibly different from those in the first step) to assess the causal effect of DNA methylation on the outcome.

# References

Adalsteinsson, B. T., Gudnason, H., Aspelund, T., Harris, T. B., Launer, L. J., Eiriksdottir, G., . . . Gudnason, V. (2012). Heterogeneity in white blood cells has potential to confound DNA methylation measurements.

Barfield, R., Feng, H., Gusev, A., Wu, L., Zheng, W., Pasaniuc, B., & Kraft, P. (2018). Transcriptome-wide association studies accounting for colocalization using Egger regression. *Genetic Epidemiology, 42*(5), 418-433.

Berger, B., Peng, J., & Singh, M. (2013). Computational solutions for omics data. *Nature Reviews Genetics, 14*(5), 333-346.

Bland, J. M., & Altman, D. G. (1995). Multiple significance tests: the Bonferroni method. *Bmj, 310*(6973), 170.

Bowden, J., Davey Smith, G., & Burgess, S. (2015). Mendelian randomization with invalid instruments: effect estimation and bias detection through Egger regression. *International Journal of Epidemiology, 44*(2), 512--525.

Bowden, J., Davey Smith, G., Haycock, P. C., & Burgess, S. (2016). Consistent estimation in Mendelian randomization with some invalid instruments using a weighted median estimator. *Genetic Epidemiology, 40*(4), 304--314.

Bowden, J., Del Greco M, F., Minelli, C., Davey Smith, G., Sheehan, N., & Thompson, J. (2017). A framework for the investigation of pleiotropy in two-sample summary data Mendelian randomization. *Statistics in Medicine, 36*(11), 1783--1802.

Brindle, J. T., Antti, H., Holmes, E., Tranter, G., Nicholson, J. K., Bethell, H. W. L., . . . Grainger, D. J. (2002). Rapid and noninvasive diagnosis of the presence and severity of coronary heart disease using 1H-NMR-based metabonomics. *Nature Medicine, 8*(12), 1439-1445. doi:10.1038/nm1202-802

Clark-Boucher, D., Zhou, X., Du, J., Liu, Y., Needham, B. L., Smith, J. A., & Mukherjee, B. (2023). Methods for mediation analysis with high-dimensional DNA methylation data: Possible choices and comparisons. *PLoS genetics, 19*(11), e1011022.

Consortium, G. P. (2010). A map of human genome variation from population scale sequencing. *Nature, 467*(7319), 1061.

Davey Smith, G., & Ebrahim, S. (2003). 'Mendelian randomization': can genetic epidemiology contribute to understanding environmental determinants of disease? *International Journal of Epidemiology, 32*(1), 1--22.

Davey Smith, G., & Hemani, G. (2014). Mendelian randomization: genetic anchors for causal inference in epidemiological studies. *Human Molecular Genetics, 23*(R1), R89--R98.

De Bakker, P. I., McVean, G., Sabeti, P. C., Miretti, M. M., Green, T., Marchini, J., . . . Delgado, M. (2006). A high-resolution HLA and SNP haplotype map for disease association studies in the extended human MHC. *Nature Genetics, 38*(10), 1166-1172.

Didelez, V., & Sheehan, N. (2007). Mendelian randomization as an instrumental variable approach to causal inference. *Statistical Methods in Medical Research, 16*(4), 309--330.

Dupont, C., Armant, D. R., & Brenner, C. A. (2009). *Epigenetics: definition, mechanisms and clinical perspective.* Paper presented at the Seminars in reproductive medicine.

Emilsson, V., Ilkov, M., Lamb, J. R., Finkel, N., Gudmundsson, E. F., Pitts, R., . . . Aspelund, T. (2018). Co-regulatory networks of human serum proteins link genetics to disease. *Science, 361*(6404), 769-773.

Enche Ady, C. N. A., Lim, S. M., Teh, L. K., Salleh, M. Z., Chin, A. V., Tan, M. P., . . . Ramasamy, K. (2017). Metabolomic-guided discovery of Alzheimer's disease biomarkers from body fluid. *Journal of neuroscience research, 95*(10), 2005-2024.

Fehrmann, R. S., Jansen, R. C., Veldink, J. H., Westra, H.-J., Arends, D., Bonder, M. J., . . . Smolonska, A. (2011). Trans-eQTLs reveal that independent genetic variants associated with a complex phenotype converge on intermediate genes, with a major role for the HLA. *PLoS genetics, 7*(8), e1002197.

Gleason, K. J., Yang, F., & Chen, L. S. (2021). A robust two-sample transcriptome-wide Mendelian randomization method integrating GWAS with multi-tissue eQTL summary statistics. *Genetic Epidemiology, 45*(4), 353-371.

Graw, S., Chappell, K., Washam, C. L., Gies, A., Bird, J., Robeson, M. S., & Byrum, S. D. (2021). Multi-omics data integration considerations and study design for biological systems and disease. *Molecular omics, 17*(2), 170-185.

Hasin, Y., Seldin, M., & Lusis, A. (2017). Multi-omics approaches to disease. *Genome biology, 18*(1), 1-15.

Hemani, G., Zheng, J., Elsworth, B., Wade, K. H., Haberland, V., Baird, D., . . . Langdon, R. (2018). The MR-Base platform supports systematic causal inference across the human phenome. *elife, 7*, e34408.

Hernandez, D. G., Nalls, M. A., Moore, M., Chong, S., Dillman, A., Trabzuni, D., . . . Weale, M. E. (2012). Integration of GWAS SNPs and tissue specific expression profiling reveal discrete eQTLs for human traits in blood and brain. *Neurobiology of disease, 47*(1), 20-28.

Houseman, E. A., Kim, S., Kelsey, K. T., & Wiencke, J. K. (2015). DNA methylation in whole blood: uses and challenges. *Current environmental health reports, 2*, 145-154.

Imming, P., Sinning, C., & Meyer, A. (2006). Drugs, their targets and the nature and number of drug targets. *Nature reviews Drug discovery, 5*(10), 821-834.

Lawlor, D. A., Harbord, R. M., Sterne, J. A., Timpson, N., & Davey Smith, G. (2008). Mendelian randomization: using genes as instruments for making causal inferences in epidemiology. *Statistics in Medicine, 27*(8), 1133-1163.

Liu, Z., Shen, J., Barfield, R., Schwartz, J., Baccarelli, A. A., & Lin, X. (2022). Large-scale hypothesis testing for causal mediation effects with applications in genome-wide epigenetic studies. *Journal of the American Statistical Association, 117*(537), 67-81.

Lord, J., Jermy, B., Green, R., Wong, A., Xu, J., Legido-Quigley, C., . . . Proitsi, P. (2021). Mendelian randomization identifies blood metabolites previously linked to midlife cognition as causal candidates in Alzheimer's disease. *Proceedings of the National Academy of Sciences, 118*(16), e2009808118.

Millwood, I. Y., Bennett, D. A., Holmes, M. V., Boxall, R., Guo, Y., Bian, Z., . . . Du, H. (2018). Association of CETP gene variants with risk for vascular and nonvascular diseases among Chinese adults. *JAMA Cardiology, 3*(1), 34-43.

Nica, A. C., Montgomery, S. B., Dimas, A. S., Stranger, B. E., Beazley, C., Barroso, I., & Dermitzakis, E. T. (2010). Candidate causal regulatory effects by integration of expression QTLs with complex trait genetic associations. *PLoS genetics, 6*(4), e1000895.

Picard, M., Scott-Boyer, M.-P., Bodein, A., Périn, O., & Droit, A. (2021). Integration strategies of multi-omics data for machine learning analysis. *Computational and Structural Biotechnology Journal, 19*, 3735-3746.

Porcu, E., Rüeger, S., Lepik, K., Santoni, F. A., Reymond, A., & Kutalik, Z. (2019). Mendelian randomization integrating GWAS and eQTL data reveals genetic determinants of complex and clinical traits. *Nature communications, 10*(1), 3300.

Purcell, S., Neale, B., Todd-Brown, K., Thomas, L., Ferreira, M. A. R., Bender, D., . . . et al. (2007). PLINK: a tool set for whole-genome association and population-based linkage analyses. *The American Journal of Human Genetics, 81*(3), 559--575.

Relton, C. L., & Davey Smith, G. (2012). Two-step epigenetic Mendelian randomization: a strategy for establishing the causal role of epigenetic processes in pathways to disease. *International Journal of Epidemiology, 41*(1), 161-176.

Sanderson, E., Glymour, M. M., Holmes, M. V., Kang, H., Morrison, J., Munafò, M. R., . . . Zhao, Q. (2022). Mendelian randomization. *Nature Reviews Methods Primers, 2*(1), 6.

Santos, R., Ursu, O., Gaulton, A., Bento, A. P., Donadi, R. S., Bologa, C. G., . . . Oprea, T. I. (2017). A comprehensive map of molecular drug targets. *Nature reviews Drug discovery, 16*(1), 19-34.

Subramanian, I., Verma, S., Kumar, S., Jere, A., & Anamika, K. (2020). Multi-omics data integration, interpretation, and its application. *Bioinformatics and biology insights, 14*, 1177932219899051.

Sun, B. B., Chiou, J., Traylor, M., Benner, C., Hsu, Y.-H., Richardson, T. G., . . . Vasquez-Grinnell, S. G. (2023). Plasma proteomic associations with genetics and health in the UK Biobank. *Nature*, 1-10.

Sun, B. B., Maranville, J. C., Peters, J. E., Stacey, D., Staley, J. R., Blackshaw, J., . . . Surendran, P. (2018). Genomic atlas of the human plasma proteome. *Nature, 558*(7708), 73-79.

Suzuki, M. M., & Bird, A. (2008). DNA methylation landscapes: provocative insights from epigenomics. *Nature Reviews Genetics, 9*(6), 465-476.

Swerdlow, D. I., Kuchenbaecker, K. B., Shah, S., Sofat, R., Holmes, M. V., White, J., . . . et al. (2016). Selecting instruments for Mendelian randomization in the wake of genome-wide association studies. *International Journal of Epidemiology, 45*(5), 1600--1616.

Tam, V., Patel, N., Turcotte, M., Bossé, Y., Paré, G., & Meyre, D. (2019). Benefits and limitations of genome-wide association studies. *Nature Reviews Genetics, 20*(8), 467-484.

Tian, P., Yao, M., Huang, T., & Liu, Z. (2022). CoxMKF: a knockoff filter for high-dimensional mediation analysis with a survival outcome in epigenetic studies. *Bioinformatics, 38*(23), 5229-5235.

Uffelmann, E., Huang, Q. Q., Munung, N. S., De Vries, J., Okada, Y., Martin, A. R., . . . Posthuma, D. (2021). Genome-wide association studies. *Nature Reviews Methods Primers, 1*(1), 59.

Vahabi, N., & Michailidis, G. (2022). Unsupervised multi-omics data integration methods: a comprehensive review. *Frontiers in Genetics, 13*, 854752.

Vandiedonck, C. (2018). Genetic association of molecular traits: a help to identify causative variants in complex diseases. *Clinical Genetics, 93*(3), 520-532.

Wang, W. Y., Barratt, B. J., Clayton, D. G., & Todd, J. A. (2005). Genome-wide association studies: theoretical and practical concerns. *Nature Reviews Genetics, 6*(2), 109-118.

Weinhold, B. (2006). Epigenetics: the science of change. In: National Institute of Environmental Health Sciences.

Westra, H.-J., Peters, M. J., Esko, T., Yaghootkar, H., Schurmann, C., Kettunen, J., . . . Powell, J. E. (2013). Systematic identification of trans eQTLs as putative drivers of known disease associations. *Nature genetics, 45*(10), 1238-1243.

Xu, S., Fung, W. K., & Liu, Z. (2021). MRCIP: a robust Mendelian randomization method accounting for correlated and idiosyncratic pleiotropy. *Briefings in Bioinformatics, 22*(5), bbab019.

Xu, S., Wang, P., Fung, W. K., & Liu, Z. (2023). A novel penalized inverse-variance weighted estimator for Mendelian randomization with applications to COVID-19 outcomes. *Biometrics, 79*(3), 2184-2195.

Yao, M., Guo, Z., & Liu, Z. (2023). Robust Mendelian Randomization Analysis by Automatically Selecting Valid Genetic Instruments for Inferring Causal Relationships between Complex Traits and Diseases. *medRxiv*, 2023.2002. 2020.23286200.

Yin, X., Chan, L. S., Bose, D., Jackson, A. U., VandeHaar, P., Locke, A. E., . . . Yu, K. (2022). Genome-wide association studies of metabolites in Finnish men identify disease-relevant loci. *Nature Communications, 13*(1), 1644.

Zhao, S., Crouse, W., Qian, S., Luo, K., Stephens, M., & He, X. (2024). Adjusting for genetic confounders in transcriptome-wide association studies improves discovery of risk genes of complex traits. *Nature Genetics*, 1-12.

Zheng, J., Haberland, V., Baird, D., Walker, V., Haycock, P. C., Hurle, M. R., . . . Luo, S. (2020). Phenome-wide Mendelian randomization mapping the influence of the plasma proteome on complex diseases. *Nature Genetics, 52*(10), 1122-1131.

Zuber, V., Colijn, J. M., Klaver, C., & Burgess, S. (2020). Selecting likely causal risk factors from high-throughput experiments using multivariable Mendelian randomization. *Nature communications, 11*(1), 29.